# Facile synthesis and step by step enhancement of blue photoluminescence from Ag-doped ZnS quantum dots

**Sonal Sahai[1,2], Mushahid Husain[2], Virendra Shanker[1], Nahar Singh[1] and D. Haranath[1,a)]**

[1]*National Physical Laboratory, Council of Scientific and Industrial Research, Dr K S Krishnan Road, New Delhi – 110 012, India*

[2]*Department of Physics, Faculty of Natural Science, Jamia Millia Islamia, New Delhi-110025, India*

[a] Corresponding author: D. HARANATH, E-mail: haranath@nplindia.org
Tel: +91-11-4560 9385, Fax: +91-11-4560 9310

**Abstract**

Our results pertaining to the step by step enhancement of photoluminescence (PL) intensity from ZnS:Ag,Al quantum dots (QDs) are presented. Initially, these QDs were synthesized using a simple co-precipitation technique involving a surfactant, polyvinylpyrrolidone (PVP), in de-ionised water. It was observed that the blue PL originated from ZnS:Ag,Al QDs was considerably weak and not suitable for any practical display application. Upon UV (365 nm) photolysis, the PL intensity augmented to ~170% and attained a saturation value after ~100 minutes of exposure. This is attributed to the photo-corrosion mechanism exerted by high-flux UV light on ZnS:Ag,Al QDs. Auxiliary enhancement of PL intensity to 250% has been evidenced by subjecting the QDs to high temperatures (200ºC) and pressures (~120 bars) in a sulphur-rich atmosphere, which is due to the improvement in crystallanity of ZnS QDs. The origin of the bright blue PL has been discussed. The results were supported by x-ray phase analysis, high-resolution electron microscopy and compositional evaluation.

## 1. Introduction

Over the past few years, scientists have been working extensively over quantum dot (QD) systems and demonstrating changes in their electrical, mechanical, optical as well as structural properties from their bulk counterparts [1]. The optical properties in QD systems is useful for various applications such as full colour display devices, white LEDs, photovoltaic cells, bio-imaging and other medical related applications [2-7]. Over the decades, zinc sulphide (ZnS), a wide and direct band gap ($E_{g\ bulk}$ = 3.6 eV) semiconductor having exciton Bohr radius, $R_B$ = 2.5 nm, has been studied to modify the color of emission using dopants such as $Mn^{2+}$ and isovalent impurities or complexes [8-12]. Bhargava et. al showed for the first time an efficient yellowish-orange emission in ZnS:$Mn^{2+}$ [12]. Afterwards, there came up many reports for generating efficient and bright orange-red (~590 nm) photoluminescence (PL) from $Mn^{2+}$ doped ZnS QDs [13-16], which is the most studied system.

Another QD system, which is moderately explored is $Cu^{+}$ doped ZnS that gives green (~530 nm) PL. The researchers have adopted several mechanisms to incorporate the Cu ion in its single valance (+1) state into the ZnS lattice so that green PL could be obtained. For example, by reducing the salts of $Cu^{2+}$ to $Cu^{+}$ state or by chelating $Cu^{+}$ ion and then incorporating it in the ZnS host lattice [17,18]. There exist reports on blue PL from ZnS QD systems also, which is mostly attributed from the surface related defect states [19-22] and/or $Ag^{+}$ doping in ZnS QDs [17]. However, the brightness levels of blue PL from ZnS QDs are weak, that limited the practical applications. On the other hand, it is known that ZnS:Ag,Al in its bulk form, is an efficient phosphor [23,24]. The current research work is an attempt to achieve stable and efficient blue PL from QDs of ZnS:Ag,Al for displays that promises better resolution, brightness and contrast.

In this paper, we report a novel method for significantly increasing the PL intensity of ZnS:Ag,Al QDs prepared by wet-chemical route. This is comprised of three consecutive processes viz. preparation of ZnS QDs by co-precipitation; UV photolysis (UVP); and UVP together with a polysulfide hydrothermal (UV+PSH) treatment. As of now, UVP treatment has been reported to improve the PL intensity for QDs such as CdSe, CdS, PbSe and ZnS:$Mn^{2+}$ [25-29]. For the first time Bhargava's group [30,31] has attempted the photon induced passivation of surface-states of ZnSMn$^{2+}$ QDs using methacrylic acid and poly(methyl methacrylate) polymer as surfactant. The process resulted in efficient surfactant adsorption and improved the luminescence efficiency of Mn$^{2+}$-doped ZnS nanocrystals. But such treatment on ZnS:Ag,Al QDs for better incorporation of Ag (dopant) has not been explored so far. Hence, we report a systematic study of UVP treatment on the PL intensity of ZnS:Ag,Al QDs as a function of UV exposure on time-base. In addition to that, we report the development of a novel polysulphide hydrothermal (PSH) method, which resulted in increased yield of ZnS QDs with improved morphological, optical and structural properties. Hydrothermal is known to be an efficient method to synthesize monodispersed nanocrystalline materials. The method is best for the synthesis of oxide nanoparticles [32-35], however, there exist reports on synthesis of sulfide nanoparticles also [36]. In preparation of sulphide nanoparticles, generally Zn:S molar ratio could not be maintained giving rise to the defects related to sulphur vacancies and sometimes formation of oxy-sulfide phases. In the present work, it is supposed that the use of the polysulfide hydrothermal method over hydrothermal could maintain the Zn:S molar ratio.

## 2. Material and Methods

To enhance the PL properties, ZnS:Ag,Al QDs were prepared using co-precipitation [12,22] method followed by UVP and UVP+PSH treatments. The detailed steps of ZnS QD preparation are illustrated in Fig. 1. The precursor chemicals such as zinc acetate [$Zn(CH_3COO)_2.2H_2O$], silver nitrate [$AgNO_3$], aluminium nitrate [$Al_2(NO_3)_3.9H_2O$] and sodium sulphide [$Na_2S$] were used as procured from Sigma Aldrich without any further purification. In a typical experiment, $AgNO_3$ and $Al_2(NO_3)_3.9H_2O$ were dissolved using de-ionised water (DIW), from Millipore. To this clear solution, calculated amount of polyvinylpyrrolidone (PVP) was added and stirred for 15 minutes, say solution 'A'. Stoichiometric amount of aluminium salt was added in order to compensate the charge conflict between $Zn^{2+}$ and $Ag^{+}$ ions. Secondly, KCl and $Zn(CH_3CHOO)_2$ were dissolved in DIW; and $Na_2S$ in DIW; say solutions 'B' and 'C', respectively. To start with, solutions 'A' and 'B' were mixed thoroughly at room temperature (~25ºC) for 15 minutes and stored up along with solution 'C' separately in a Julabo F-25 chilling bath maintained at 5±0.1ºC. After an hour of chilling, solution 'C' was added drop-wise to the above mixture of solutions under continuous stirring. Slight turbidity (white) appeared in the solution after 25 minutes, which indicates the formation of ZnS QDs. This colloidal solution was further divided into three parts and subjected to ambient conditions, UVP and UVP+PSH treatments, respectively. For UVP, the samples were irradiated for ~180 minutes by 125 W high flux UV lamp having wavelength of 365 nm, whereas, for PSH treatment, the UV irradiated sample was cooked with a polysulphide solution in a high temperature-pressure autoclave for 2 hours. The ratio of silver to zinc mentioned in Fig. 1 is the optimized value. However, there observed a black precipitate of $Ag_2S$ formation instead of Ag doped ZnS, when we increase the ratio >2%.

For x-ray diffraction (XRD) analysis, QD agglomerates were precipitated out from solution by adding sufficient amount of ethanol and aging them for one week in a dehumidifier chamber. To determine the phase purity of the ZnS QDs, the crystalline data was obtained by x-ray diffractometer (XRD; Bruker-AXS D8 Advance system using Cu $K_\alpha$ ($\lambda$ ~1.54056 Å) with step size of 0.01 at 0.15 steps/sec. High-resolution transmission electron microscopy (HRTEM) was used to record the images of ZnS QDs with FEI (Technai G2 20 S-Twin) transmission electron microscope operating at 200 kV. The room-temperature photoluminescence (PL) spectra were recorded at regular intervals of time by PerkinElmer LS-55 spectrophotometer keeping the instrument settings constant for all the samples at the scan speed of 300 nm/min. Time-resolved photoluminescence (TRPL) spectrum was recorded using a microsecond xenon flash lamp and a nanosecond hydrogen flash lamp, respectively as the source of excitations, attached to luminescence spectrometer (Edinburgh F-900). Unfortunately, we could not observe any shorter relaxation processes that are in the nano or pico range with the use of nanosecond hydrogen flash lamp.

## 3. Results and Discussion

### *3.1. UV photolysis*

UV photolysis (UVP) is a versatile phenomenon of increasing PL intensity of QDs by the irradiation of high flux UV light. This is an important step to enhance the luminescence and has been observed in many of the QD systems such as CdS, CdSe, PbSe etc [25-29]. This increase of PL has also been named as "photo-activation". The photo-activation of QDs was first time reported by Cordero *et al.* [25] in CdSe nanoparticles. ZnS based QD systems were less explored for increase in PL intensity by UV photolysis. However, there are few reports where increase in PL has been observed, when ZnS:$Mn^{2+}$ nanoparticles were irradiated by UV light [13,30]. Here, we report a detailed study on increase in PL intensity of ZnS:Ag,Al QDs by UVP mechanism using high flux (125 W) and monochromatic (365 nm) UV lamp. The effect of UVP process on PL enhancement of ZnS:Ag,Al QDs is shown in Fig. 2. It has been observed that PL intensity increases with UV irradiation time and gets saturated after ~100 minutes. The reason for PL enhancement in ZnS:Ag,Al samples by UVP could be understood as the adsorption of water molecules on the surface of QDs during the early illumination times and photo-corrosion due to oxygen present in the aqueous solution. Initial increase in PL emission may be due to adsorption of water molecules, which is supposed to passivate the surface charge carrier traps. The presence of oxygen induces a slow photo-corrosion, resulting in the release of $Zn^{2+}$ according to the following reaction, which is very similar to the reaction suggested for CdS QDs [37, 38].

$$ZnS + 2H_2O + O_2 \longrightarrow Zn^{2+} + SO_4^{2-} + 4H^+$$

Immediately after the reaction, PL intensity significantly increases due to formation of Zn-OH or $Zn(OH)_2$ on the surface of ZnS. The formation of Zn-OH bonds or, hydroxide layer eliminates the surface defects, thereby enhancing the PL intensity [39].

*3.2. Structure and Morphology*

X-ray diffraction (XRD) patterns of the ZnS QDs after UVP and UVP+PSH treatments were shown in Fig. 3(a) and (b), respectively. Both the samples did show zinc blende (sphalerite) structure. It could be seen from Fig. 3(a) that the ZnS:Ag,Al QDs treated by UVP process, show only two peaks, (111) and (200) that shows poor crystallanity of the sample. Whereas, the sample treated with UVP+PSH process showed improved crystallanity [Fig. 3(b)], which is evidenced from the well defined peaks at (111), (200), (220) and (311) of the XRD pattern. Both the XRD patterns match very well with the JCPDS card no. 05-0566 and confirms the phase purity of ZnS QDs.

Low and high resolution transmission electron micrographs for the UVP+PSH treated ZnS QDs are shown in Fig. 4(a) and (b), respectively. It could be seen from Fig. 4(a) that the nanoparticles are well separated and more or less spherical. The inset of Fig. 4(a) depicts the electron diffraction image of ZnS QDs, consisting of three concentric sharp rings, which correspond to (111), (220) and (311) diffraction of the cubic structure. Clear lattice fringes are observed for ZnS QDs in the Fig. 4(b). Gatan Digital Micrograph software was used to determine the fringe width to be 0.32 nm, which is in good agreement with the d spacing value of the plane (111) of ZnS QDs. The same has been shown in the inset of Fig. 4(b).

From the electron microscopy analysis the diameter of the particles was found to be ~5 nm, which tells that the particles are in quantum confinement region ($R_B$ = 2.5 nm) and thus they exhibit improved luminescent properties, as described in the later section. The chemical composition of ZnS QDs was examined by energy dispersive x-ray spectrum (EDX) data. Fig. 5(a) and (b) shows the EDX spectra of ZnS QDs treated by UVP and UVP + PSH processes, respectively. Zn and S signals in both the spectra corresponds to ZnS, while C and Cu element signals observed are due to the carbon coated copper grid. The strength of Zn and S signals is stronger in case of UVP + PSH treated samples, which confirms the formation of ZnS QDs with better yield. As the concentrations of $Ag^+$ (dopant) and charge compensating $Al^{3+}$ (co-dopant) ions used were 0.4 mM each, the EDX instrument could not detect the signals related to $Ag^+$ and/or $Al^{3+}$ ions. However, we confirmed their presence as Ag = 25±2 ppm and Al = 8±2 ppm, in the sample using atomic absorption spectrometer, which is relatively sensitive than EDX machine.

*3.3. Photoluminescence (PL) and Time-resolved photoluminescence (TRPL)*

Fig. 6 compares the room-temperature photoluminescence spectra of the ZnS QDs prepared by co-precipitation, UVP and UVP+PSH treatments, respectively. Enhancement of PL in respective curves is clearly observed at various stages. The UVP+ PSH treated ZnS QDs show enhancement of their PL up to 250% compared to the sample made by co-precipitation technique, which is a significant achievement. It is important to note that ZnS QDs prepared by co-precipitation exhibits negligible PL, peaking at ~420 nm. Upon UVP process, the PL intensity has increased to ~170% and attained saturation after ~100 minutes with no change in the PL peak position, whereas, after UVP+PSH treatment, there was a red-shift in PL peak position by ~36 nm with a remarkable enhancement (250%) in PL intensity. The reason for the improvement in PL brightness is attributed to efficient donor-acceptor pair recombinations. The PL peak observed at ~420 nm is arising due to the self-activated (SA) luminescent center by the spatial association of $Zn^{2+}$ vacancy with a co-activator $Al^{3+}$ at the nearest-neighbour site [24]. The PL peak at 456 nm could be attributed to blue-silver (B-Ag) center formed by activator as acceptor i.e. $Ag^{+}$ and co-activator as donor i.e. $Al^{3+}$. When activator and co-activators are introduced with nearly equal concentrations, they are distributed randomly in the lattice occupying the appropriate lattice sites. PL excitation peak shows a blue shift from 336 nm for QDs prepared by co-precipitation to 320 nm for QDs after UVP process. This blue shift could be attributed to a secondary effect of photo-corrosion process, which is in other words attributed as reduction in diameter of the ZnS QDs [36]. However, red shift in PL excitation peak position (~336 nm) could be due to the slight increase in size of QDs, which in turn is responsible for the presence of high temperature (200°C) and pressure (~120 bars) inside the autoclave.

Time resolved photoluminescence (TRPL) is a non-destructive and powerful technique commonly used for optical characterization of semiconductors. PL measurements done in the present study inferred that UV+PSH is one of the best mechanisms to prepare ZnS:Ag,Al QDs with exceptional PL brightness levels. Therefore, TRPL measurement was performed for this sample only. TRPL spectrum was recorded at 456 nm and 336 nm for emission and excitation, respectively. Fig. 7 gives the information about the traps associated with the material and delay in radiative recombination of ZnS:Ag,Al QDs. The plot in Fig. 7 is the best represented by the expression: decay = $A+B_1\exp(-t/\tau_1) + B_2\exp(-t/\tau_2)$, where, $\tau_1$ and $\tau_2$ are the decay constants; and A, $B_1$ and $B_2$ are constants related to the equation. The two observed decay constants are associated with fast and slow decay components having the value s 1.72 μs ($\tau_1$) and 53.07 μs ($\tau_2$) respectively. The two decay components are as a result of the two distinct radiative events. First, the fast decay could be ascribed to the direct radiative transitions of the excitons from donor to acceptor level while the second, the slower component should be due to the radiative recombination via surface-trap sites. Finally, it could be concluded that the shortest radiative lifetimes (~1.7 μs) observed for UVP+PSH treated ZnS:Ag,Al QD samples are the best choice for blue component in nano-based display systems, where the fast response is a prerequisite to avoid image overlapping or ghost-image formation.

## 4. Conclusions

The steps involved in enhancing the blue photoluminescence from ZnS:Ag,Al QDs synthesized using co-precipitation technique was systematically investigated. The size of QDs was determined to be ~5nm, which shows that they are in quantum confined regime. It was demonstrated that UV (365 nm) photolysis improved the PL intensity to 170% as that of co-precipitated sample; whereas UVP+PSH treatment appreciably improved it to 250%. The highest possible PL brightness for the blue region is obtained from UVP+PSH treated QD samples. Finally, the TRPL spectrum of these highly luminescent QDs was measured, and concluded that the samples with the shortest radiative decay time (~1.7 μs) would be the best choice for their use in nanophosphor based display devices, so that there is no lag in image transformation.

## Acknowledgments

The authors (SS and DH) wish to gratefully acknowledge the help rendered by Prof B R Mehta of IIT, Delhi, India, in getting TEM images of the samples and the Department of Science and Technology (DST) for the financial support to carry out the above research work.

**Figure captions:**

**Figure 1.** Flow-chart describing preparation of ZnS:Ag,Al QD by co-precipitation followed by UV photolysis (UVP) and polysulfide hydrothermal (PSH) treatments.

**Figure 2.** Variation of PL intensity with UV (365 nm) irradiation time (dots) and least square fitted polynomial curve (solid line)

**Figure 3.** XRD patterns of ZnS:Ag,Al QDs (a) after UVP and (b) UVP+PSH treatments.

**Figure 4.** Low and high resolution TEM micrographs of ZnS:Ag,Al QDs. The left and right insets show the diffraction pattern and magnified image of the encircled fringes respectively.

**Figure 5.** EDX analysis performed on ZnS:Ag,Al QDs (a) after UVP and (b) UVP+PSH treatments.

**Figure 6.** Room-temperature PL and PLE spectra of ZnS:Ag,Al QDs made by co-precipitation (o), after UVP (□) and UVP+PSH (▲) treatments. For recording PL and PLE, the excitation and emission wavelengths were fixed at 336 and 456 nm, respectively. Inset shows the photographs of glass vials with stages of enhanced PL, when subjected to UVP and UVP+PSH treatments.

**Figure 7.** The TRPL spectrum of ZnS:Ag,Al QDs after UVP+PSH treatment. The plot follows an exponential decay as per the equation mentioned in the figure along with the derived decay constants. The insets show the fast and slow decay components with their least square fit.

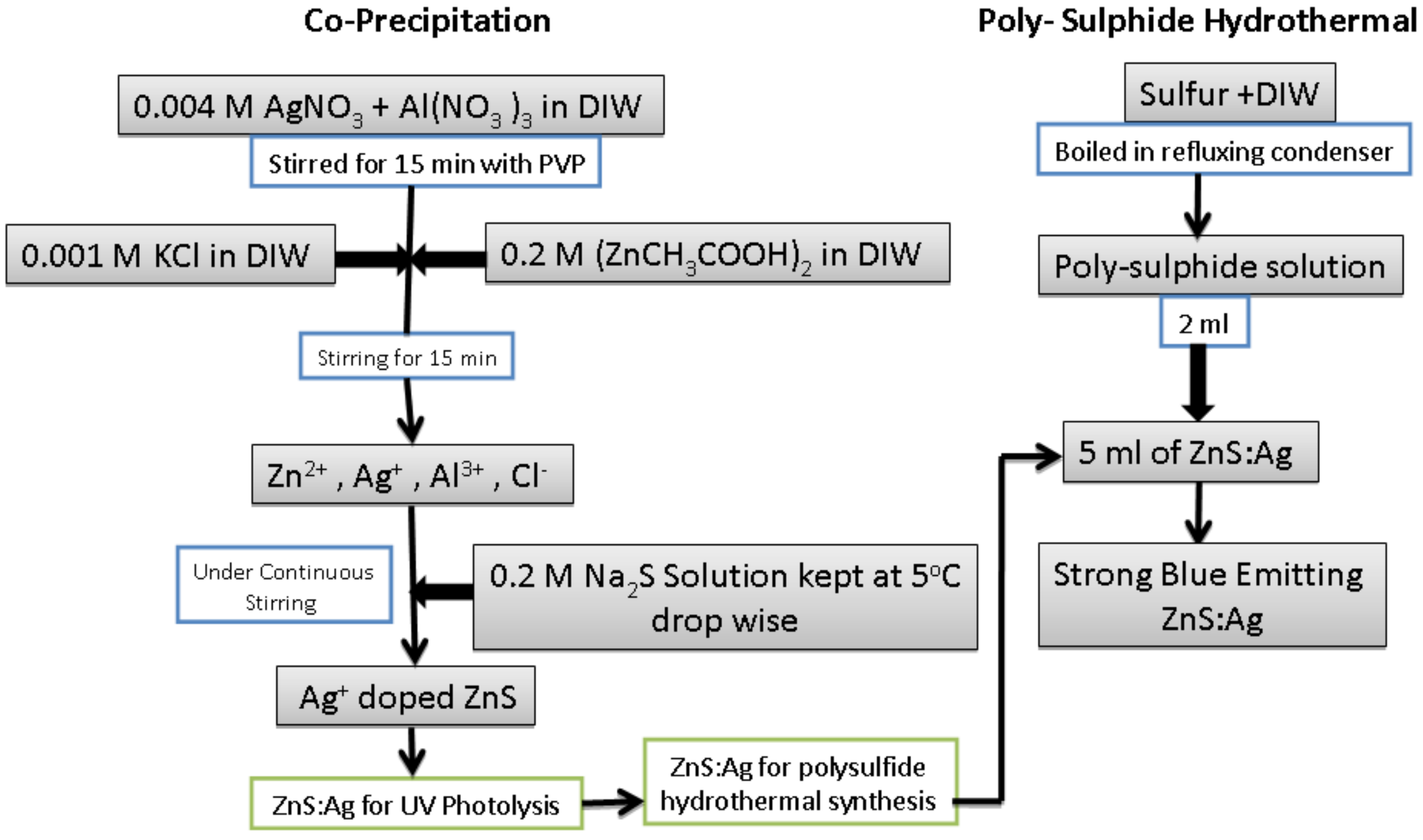

**Figure 1**

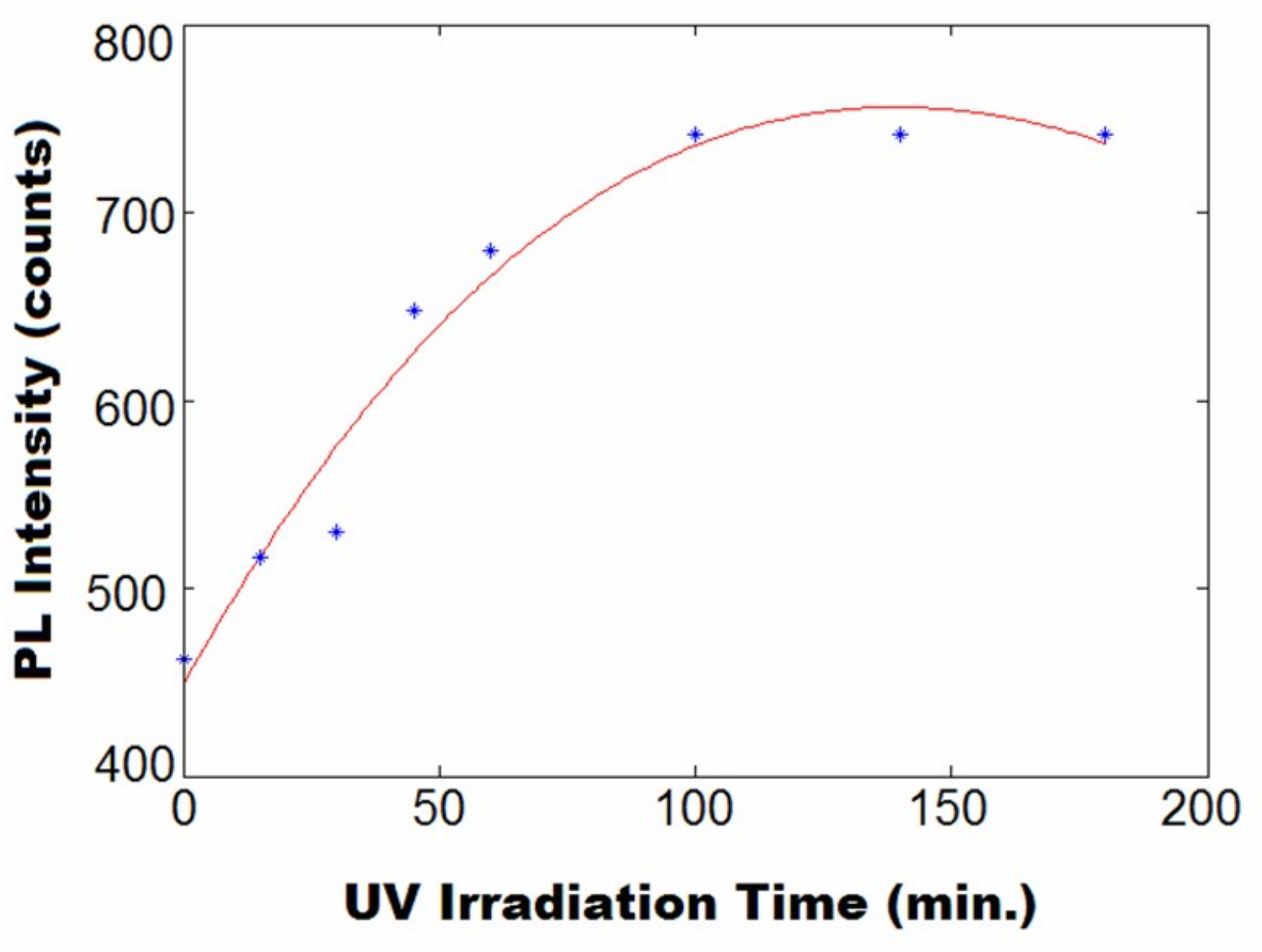

**Figure 2**

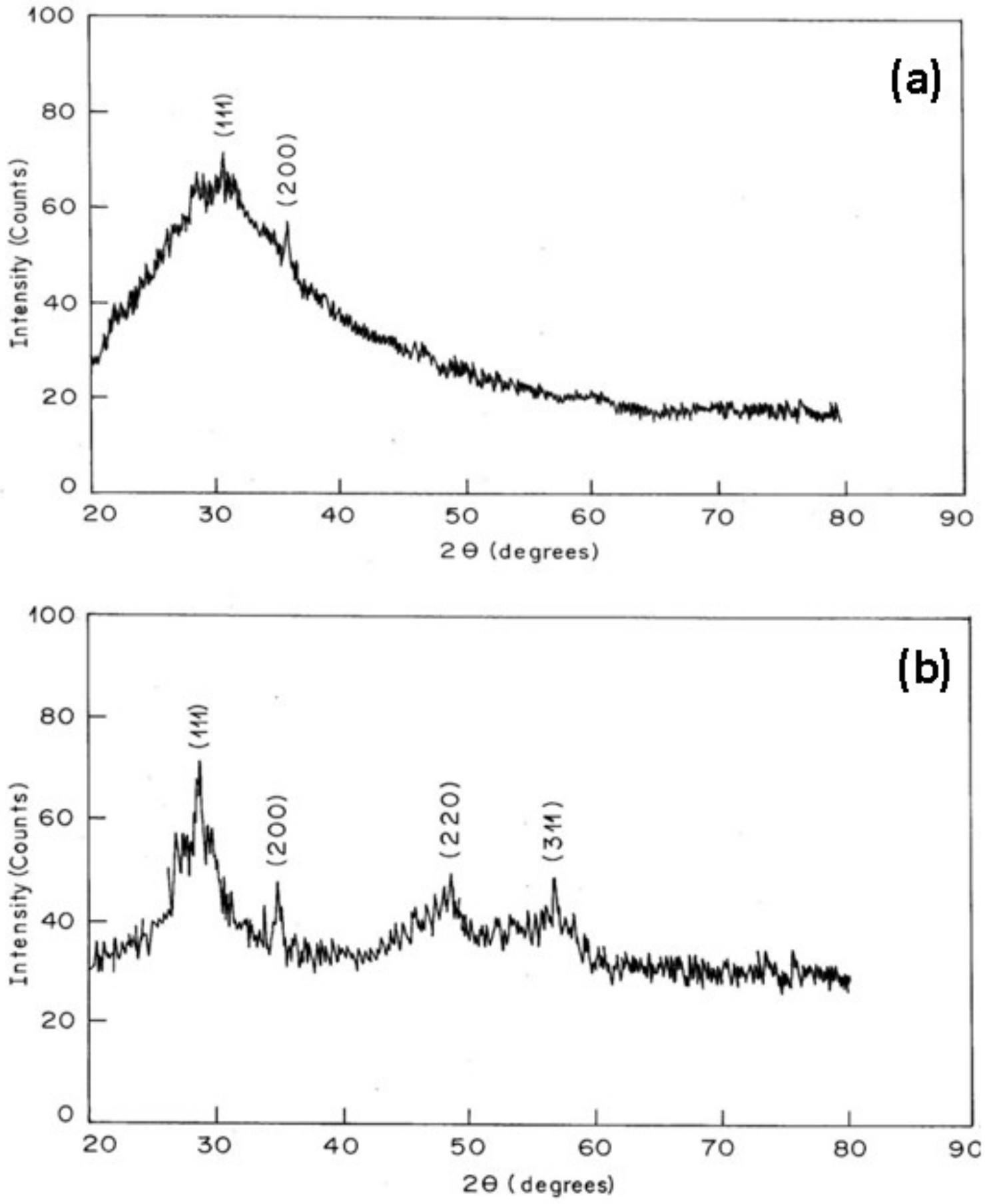

**Figure 3**

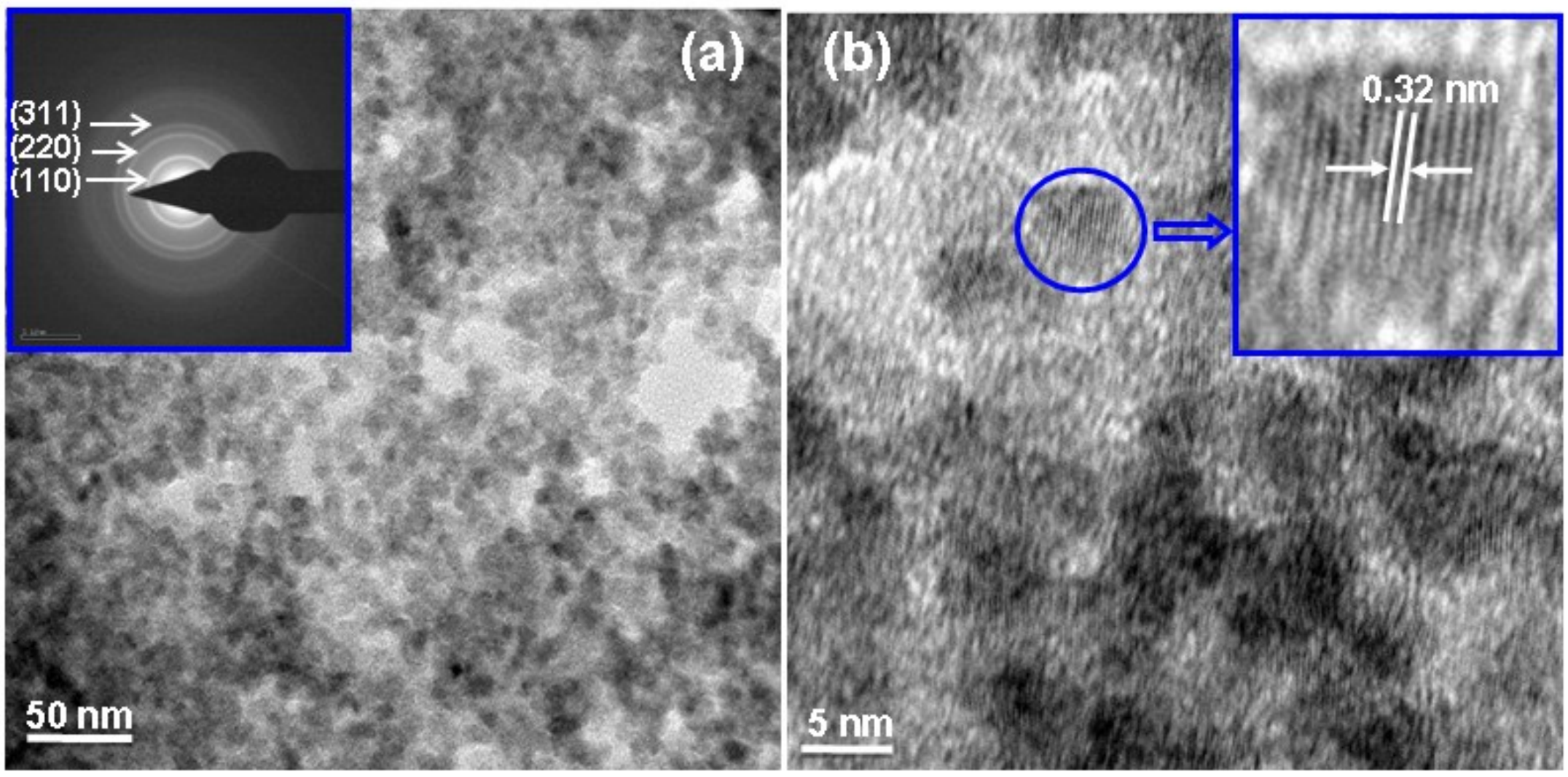

**Figure 4**

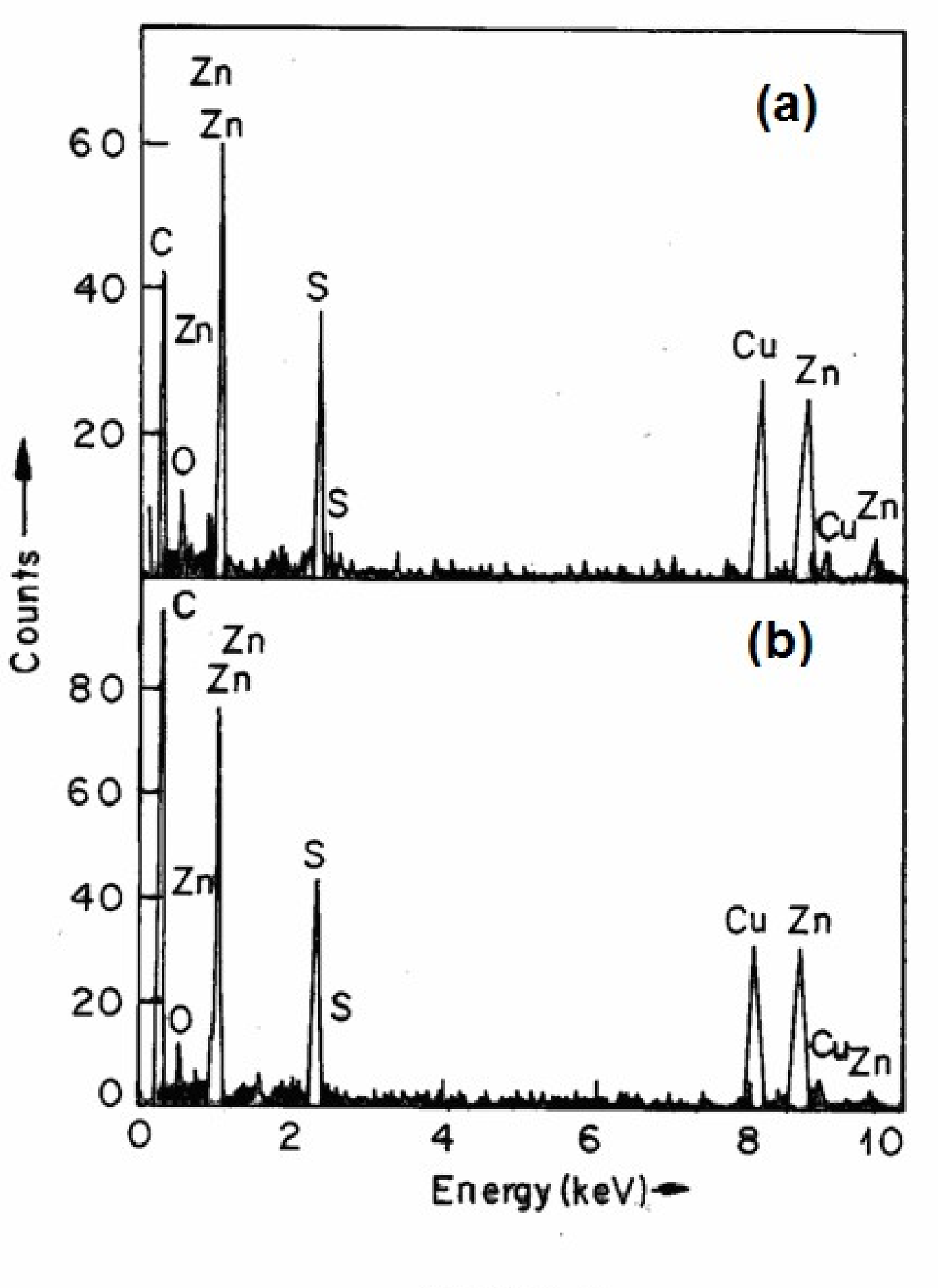

Figure 5

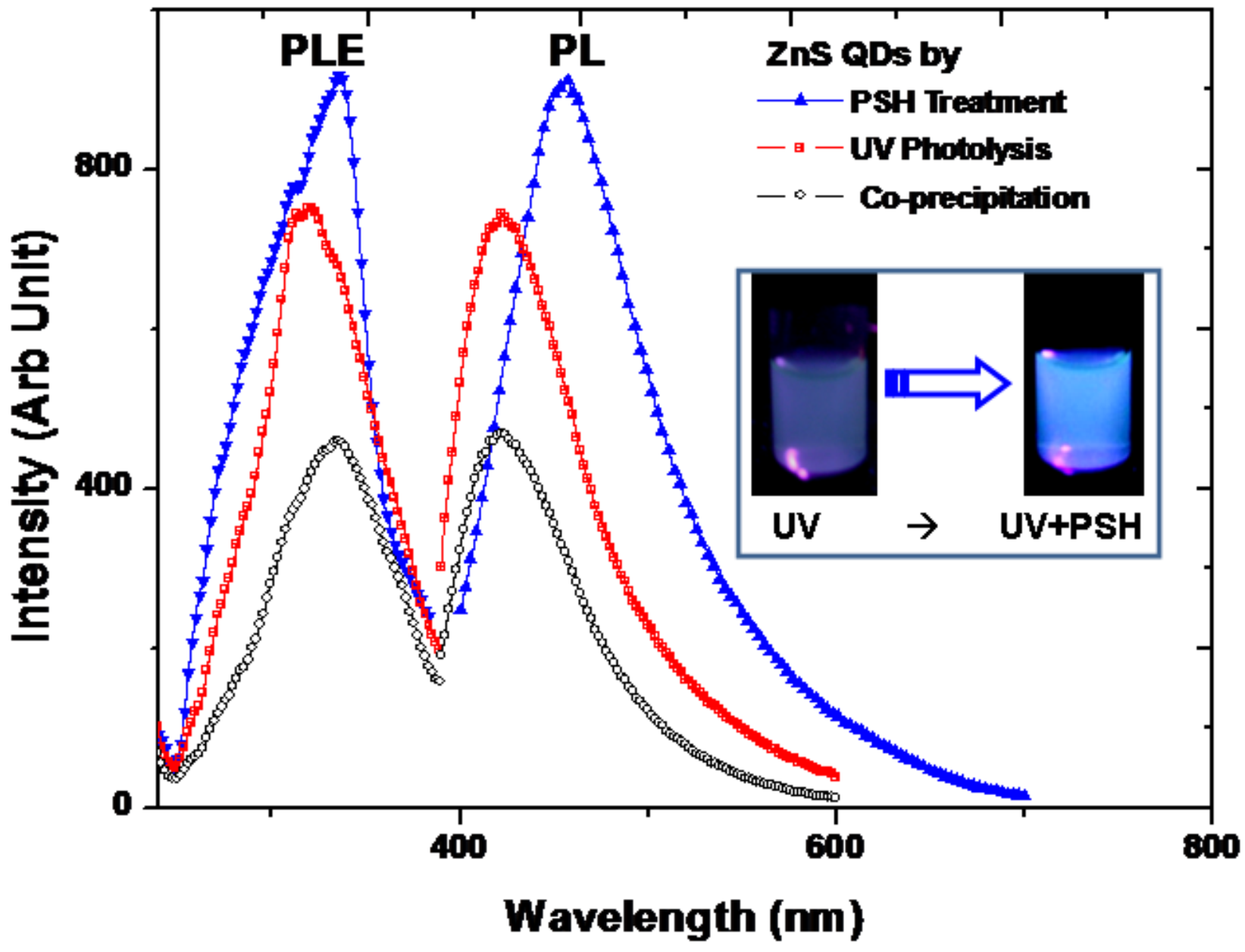

Figure 6

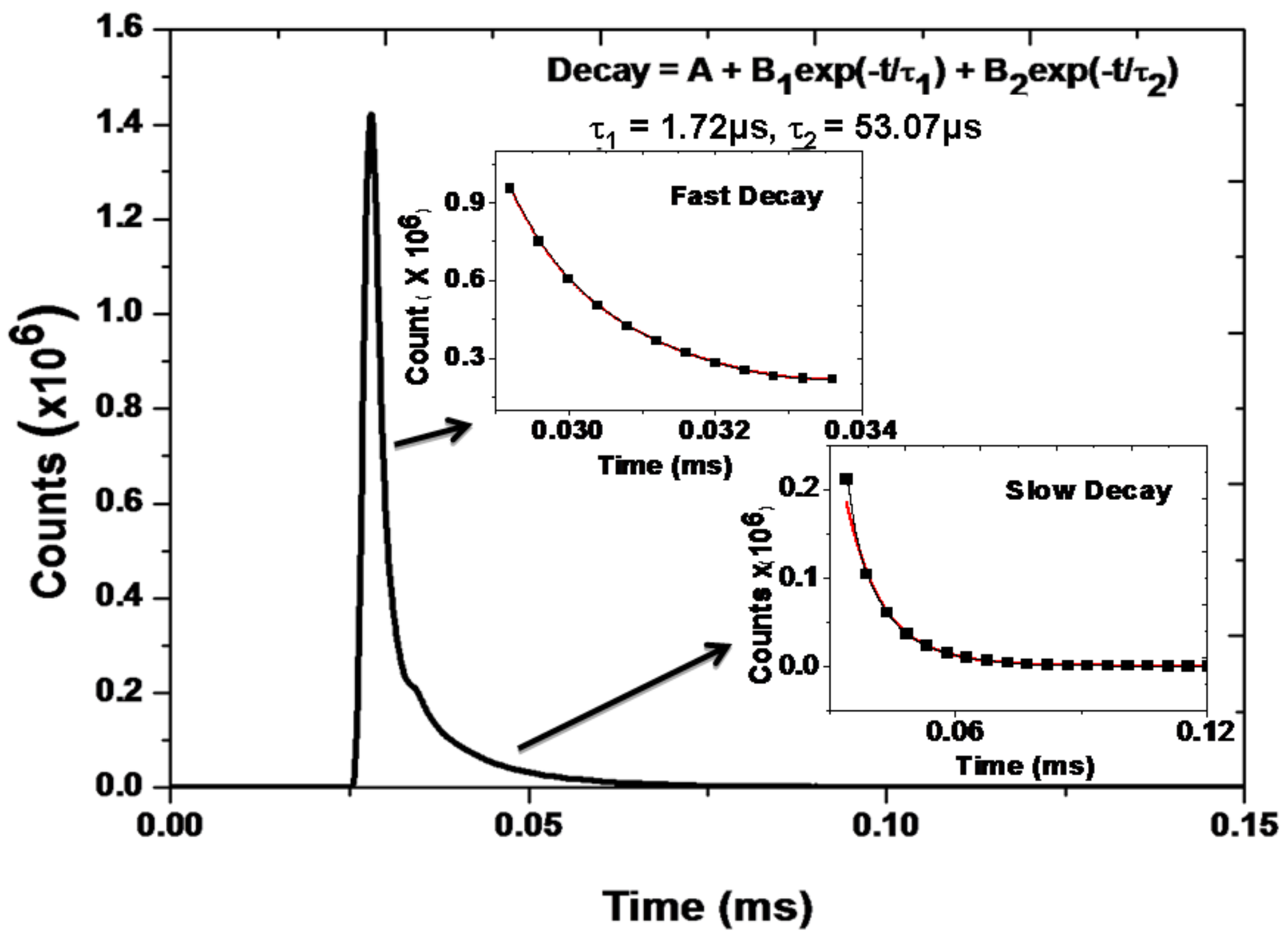

Figure 7